\documentclass[%
reprint,
nofootinbib,
 amsmath,amssymb,
 author-year
prb,
showkeys
]{revtex4-1}

\usepackage{graphicx}
\usepackage{dcolumn}
\usepackage{bm}
\usepackage{tikz} 

\begin{document}

\title{Efficient Tree Decomposition of High-Rank Tensors}

\author{Adam S. Jermyn}
\email{adamjermyn@gmail.com}
\affiliation{Institute of Astronomy, University of Cambridge, Madingley Rise, CB3 0HA, UK}

\date{\today}

\begin{abstract}

Tensors are a natural way to express correlations among many physical variables, but storing tensors in a computer naively requires memory which scales exponentially in the rank of the tensor.
This is not optimal, as the required memory is actually set not by the rank but by the mutual information amongst the variables in question.
Representations such as the tensor tree perform near-optimally when the tree decomposition is chosen to reflect the correlation structure in question, but making such a choice is non-trivial and good heuristics remain highly context-specific.
In this work I present two new algorithms for choosing efficient tree decompositions, independent of the physical context of the tensor.
The first is a brute-force algorithm which produces optimal decompositions up to truncation error but is generally impractical for high-rank tensors, as the number of possible choices grows exponentially in rank.
The second is a greedy algorithm, and while it is not optimal it performs extremely well in numerical experiments while having runtime which makes it practical even for tensors of very high rank.

\end{abstract}

\keywords{Tensor Networks; Singular Value Decomposition}
\maketitle


\section{Introduction}

In recent years the language of tensors and tensor networks has become the standard for discussing correlations amongst physical variables\ \citep{ORUS2014117}.
Disparate concepts such as lattice partition functions\ \citep{PhysRevLett.103.160601}, differential equations\ \citep{Bachmayr:2016:TNH:3027165.3027175}, high-energy field theories\ \citep{Evenbly2011}, and quantum computing\ \citep{doi:10.1137/050644756} are now being written and examined in this language, and this trend is only increasing.

Alongside the conceptual shift towards tensors and tensor networks has come a numerical shift.
Techniques such as the density matrix renormalization group (DMRG) and matrix product states (MPS) are now recognized as special cases of tensor network manipulations and representations\ \citep{ORUS2014117}.
Likewise there have been renewed efforts towards multidimensional numerical renormalization, and these lean heavily on fundamental tensor operations\ \citep{PhysRevLett.118.250602, PhysRevB.95.045117, PhysRevLett.103.160601, PhysRevLett.118.110504}.

Along the way a few key operations have been identified as bottlenecks\ \citep{ORUS2014117, PhysRevB.89.245118}.
Among the more prominent are contraction of networks\ \citep{PhysRevLett.118.110504, PhysRevE.90.033315} and efficient representation of individual tensors\ \citep{doi:10.1137/07070111X}.
This work is focused on the latter.

There have been several attempts to devise efficient means of representing high-rank tensors\ \citep{doi:10.1137/07070111X, 2016arXiv160700050Y, PhysRevLett.101.110501}, and among the most successful for physical applications is the tree tensor\ \citep{doi:10.1063/1.4798639}. Tree tensors represent a means of factoring a high-rank tensor into an acyclic network (tree) formed of rank-3 tensors\ \citep{Hackbusch2009}.
The advantage of the tree representation is that it may be contracted efficiently: matrix elements  of tree networks may be computed with little memory overhead.

Factoring alone does not produce a more compact representation.
To reduce the storage requirements of a tensor tree, the dimension of the bonds between constituent tensors must be restricted, either globally or using an error cutoff\ \citep{Hackbusch2009}.
In the latter case the network represents an approximation of the original tensor, and if the underlying correlations are weak or well-localized then the approximation may be made extremely good even with low-dimensional bonds.

The main outstanding difficulty with using tensor trees for this purpose lies in picking a decomposition\ \citep{BALLANI2013639}: Which variables or indices ought to go at which points on the tree?
Historically this has been answered heuristically, and said to depend on the context of the problem at hand\ \citep[see e.g.][]{doi:10.1063/1.4798639}.
For instance, if there is a physical reason to expect two sets of variables to be weakly coupled, they can be placed on opposing sides of the tree.
More recently methods which automate the selection have been proposed~\citep{Ballani2014} and found to perform well.
The purpose of this work is to provide two new algorithms in this vein.

In Section\ \ref{sec:tree} I review the construction of tensor trees.
I then outline a brute force algorithm in Section\ \ref{sec:brute} and a much faster greedy algorithm in Section\ \ref{sec:algo}.
Both algorithms choose decompositions based purely on the correlation properties of the tensor, with no additional physical reasoning required.
In Section\ \ref{sec:numerics} I examine the results of numerical experiments and benchmarks of both algorithms.
The brute-force algorithm provides an optimal result up to truncation error.
Unfortunately it is too slow to be practical for high-rank tensors.
The greedy algorithm is practical even at very high rank while producing results which are almost as good, and so is generally much more useful.

The algorithms and experiments discussed in this work are provided in an open-source Python package.
Details of this software are given in Appendix\ \ref{appen:software}.

\section{Tensor Trees}
\label{sec:tree}

\subsection{Singular Value Decomposition}

\begin{figure}
	\begin{tikzpicture}
	    \node[draw=none] (v0) at (-5,0) {};
	    \node[draw=none] (v1) at (-3,0) {};
	    \node[rectangle,minimum width = 3em, minimum height = 3em, draw] (v4) at (-4,0) {$M$};
	    \draw [thick]
	    (v0) -- (v0 -|  v4.west) 
	    (v1) -- (v1 -| v4.east); 

	    \node[draw=none] (u0) at (-2.5,0) {};
	    \node[draw=none] (u1) at (2.5,0) {};
	    \node[rectangle,minimum width = 3em, minimum height = 3em, draw] (u2) at (-1.5,0) {$U$};
	    \node[rectangle,minimum width = 3em, minimum height = 3em, draw] (u3) at (0,0) {$\Sigma$};
	    \node[rectangle,minimum width = 3em, minimum height = 3em, draw] (u4) at (1.5,0) {$V^\dagger$};
	    \draw [thick]
	    (u0) -- (u0 -|  u2.west) 
	    (u2.east) -- (u2.east -| u3.west)
	    (u3.east) -- (u3.east -| u4.west)
	    (u1) -- (u1 -| u4.east); 

		\draw[line width=0.3mm, transform canvas={yshift=-1.5pt}] (v1) -- (u0);
		\draw[line width=0.3mm, transform canvas={yshift=1.5pt}] (u0) -- (v1);

	\end{tikzpicture}
	\caption{The decomposition of a matrix $M$ into a product of two unitary matrices and a diagonal matrix with non-negative real entries. Squares represent tensors and lines represent indices.
Where lines attached to different tensors connect those indices are to be contracted.}
	\label{fig:svd}
\end{figure}
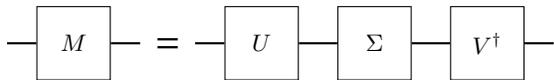

The fundamental building block of a tensor tree is the singular value decomposition (SVD), or equivalently the Schmidt decomposition\ \citep{doi:10.1142/S0129183111016683}.
This decomposition allows us to write any matrix $M$ of shape $(n,m)$ in the form
\begin{equation}
	M = U \Sigma V^\dagger,
\label{eq:svd}
\end{equation}
where $U$ and $V$ are $(n,k)$ and $(k,m)$ unitary matrices, $k=\min(n,m)$ and $\Sigma$ is a $(k,k)$ diagonal matrix with non-negative real entries.
The diagonal entries of $\Sigma$ are known as the singular values of $M$.
This decomposition is shown diagrammatically in Penrose notation\ \citep{Penrose} in Fig.\ \ref{fig:svd}.
Briefly, squares represent tensors and lines represent indices.
Where lines attached to different tensors connect those indices are to be contracted.

If $M$ has matrix rank less than $k$ some diagonal entries of $\Sigma$ vanish, and in this case the corresponding columns in $U$ and rows in $V^\dagger$ may be eliminated.
In this way the SVD allows us to produce a compressed representation of $M$.
Similarly, if several diagonal entries of $\Sigma$ are smaller than some threshold $\epsilon$ we may eliminate them and write
\begin{equation}
	M \approx \bar{U} \bar{\Sigma} \bar{V}^\dagger,
\label{eq:approxSVD}
\end{equation}
where the bar indicates that these matrices have had entries eliminated.
This approximation is optimal in the sense that there is no better low rank linear approximation for a given threshold\ \citep{Eckart1936}.

As a simple example, consider the matrix
\begin{align}
M = \frac{1}{\sqrt{2}}
\left(
\begin{array}{cc}
1 & 1\\
1 & 1
\end{array}
\right).
\label{eq:Msimp}
\end{align}
This may be decomposed as
\begin{align}
U &= (1,1),\\
V &= (1,1)\\
\intertext{and}
\Sigma &= (1).
\end{align}
If we perturb this matrix to
\begin{align}
M =
\frac{1}{\sqrt{2}}
\left(
\begin{array}{cc}
1 & 1\\
1 & 1
\end{array}
\right)
+ \frac{\delta}{\sqrt{2}}
\left(
\begin{array}{cc}
1 & -1\\
-1 & 1
\end{array}
\right),
\end{align}
then the full decomposition has
\begin{align}
U &= \frac{1}{\sqrt{2}}
\left(
\begin{array}{cc}
1 & 1\\
1 & -1
\end{array}
\right),\\
V &= \frac{1}{\sqrt{2}}
\left(
\begin{array}{cc}
1 & 1\\
1 & -1
\end{array}
\right)\\
\intertext{and}
\Sigma &= \frac{1}{\sqrt{2}}
\left(
\begin{array}{cc}
1 & 0\\
0 & \delta
\end{array}
\right).
\end{align}
If $\delta < \epsilon$ we may choose to truncate this decomposition by eliminating the second row and column in $\Sigma$, producing the same decomposition as that of equation\ \eqref{eq:Msimp}.

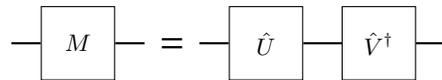
\begin{figure}
	\begin{tikzpicture}
	    \node[draw=none] (v0) at (-5,0) {};
	    \node[draw=none] (v1) at (-3,0) {};
	    \node[rectangle,minimum width = 3em, minimum height = 3em, draw] (v4) at (-4,0) {$M$};
	    \draw [thick]
	    (v0) -- (v0 -|  v4.west) 
	    (v1) -- (v1 -| v4.east); 

	    \node[draw=none] (u0) at (-2.5,0) {};
	    \node[draw=none] (u1) at (1,0) {};
	    \node[rectangle,minimum width = 3em, minimum height = 3em, draw] (u2) at (-1.5,0) {$\hat{U}$};
	    \node[rectangle,minimum width = 3em, minimum height = 3em, draw] (u4) at (0,0) {$\hat{V}^\dagger$};
	    \draw [thick]
	    (u0) -- (u0 -|  u2.west) 
	    (u2.east) -- (u2.east -| u4.west)
	    (u1) -- (u1 -| u4.east); 

		\draw[line width=0.3mm, transform canvas={yshift=-1.5pt}] (v1) -- (u0);
		\draw[line width=0.3mm, transform canvas={yshift=1.5pt}] (u0) -- (v1);

	\end{tikzpicture}
	\caption{The decomposition of a matrix $M$ into a product of two matrices. Squares represent tensors and lines represent indices.
Where lines attached to different tensors connect those indices are to be contracted.}
	\label{fig:svdp}
\end{figure}

As a notational convenience, after truncating $\Sigma$ we define
\begin{align}
	\hat{U} \equiv U \sqrt{\Sigma}\\
\intertext{and}
	\hat{V} \equiv V \sqrt{\Sigma},
\end{align}
and likewise for the approximate matrices.
Note that the matrix square root is defined for $\Sigma$ because it is diagonal.
With this notation equation\ \eqref{eq:svd} becomes
\begin{equation}
	M = \hat{U} \hat{V}^\dagger
\end{equation}
and likewise for equation\ \eqref{eq:approxSVD}.
Fig.\ \ref{fig:svd} then becomes Fig.\ \ref{fig:svdp}.

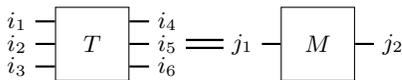
\begin{figure}
	\begin{tikzpicture}
	    \node[draw=none] (v0) at (-3,0.3) {$i_1$};
	    \node[draw=none] (v1) at (-3,0) {$i_2$};
	    \node[draw=none] (v2) at (-3,-0.3) {$i_3$};
	    \node[draw=none] (u0) at (-1,0.3) {$i_4$};
	    \node[draw=none] (u1) at (-1,0) {$i_5$};
	    \node[draw=none] (u2) at (-1,-0.3) {$i_6$};
	    \node[rectangle,minimum width = 3em, minimum height = 3em, draw] (v4) at (-2,0) {$T$};
	    \draw [thick]
	    (v0) -- (v0 -|  v4.west) 
	    (v1) -- (v1 -|  v4.west) 
	    (v2) -- (v2 -|  v4.west) 
	    (u0) -- (u0 -|  v4.east) 
	    (u1) -- (u1 -|  v4.east) 
	    (u2) -- (u2 -|  v4.east); 

	    \node[draw=none] (t0) at (0,0) {$j_1$};
	    \node[draw=none] (t1) at (2,0) {$j_2$};
	    \node[rectangle,minimum width = 3em, minimum height = 3em, draw] (t2) at (1,0) {$M$};
	    \draw [thick]
	    (t0) -- (t0 -|  t2.west) 
	    (t1) -- (t1 -| t2.east); 

		\draw[line width=0.3mm, transform canvas={yshift=-1.5pt}] (t0) -- (u1);
		\draw[line width=0.3mm, transform canvas={yshift=1.5pt}] (u1) -- (t0);

	\end{tikzpicture}
	\caption{The flattening of a tensor $T$ into a matrix $M$. Note that $j_1$ and $j_2$ are composite indices representing $\{i_1,i_2,i_3\}$ and $\{i_4,i_5,i_6\}$ respectively.}
	\label{fig:flatten}
\end{figure}

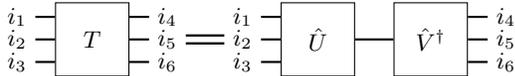
\begin{figure}
	\begin{tikzpicture}
	    \node[draw=none] (v0) at (-3,0.3) {$i_1$};
	    \node[draw=none] (v1) at (-3,0) {$i_2$};
	    \node[draw=none] (v2) at (-3,-0.3) {$i_3$};
	    \node[draw=none] (u0) at (-1,0.3) {$i_4$};
	    \node[draw=none] (u1) at (-1,0) {$i_5$};
	    \node[draw=none] (u2) at (-1,-0.3) {$i_6$};
	    \node[rectangle,minimum width = 3em, minimum height = 3em, draw] (v4) at (-2,0) {$T$};
	    \draw [thick]
	    (v0) -- (v0 -|  v4.west) 
	    (v1) -- (v1 -|  v4.west) 
	    (v2) -- (v2 -|  v4.west) 
	    (u0) -- (u0 -|  v4.east) 
	    (u1) -- (u1 -|  v4.east) 
	    (u2) -- (u2 -|  v4.east); 

	    \node[draw=none] (t0) at (0,0.3) {$i_1$};
	    \node[draw=none] (t1) at (0,0) {$i_2$};
	    \node[draw=none] (t2) at (0,-0.3) {$i_3$};
	    \node[draw=none] (q0) at (3.5,0.3) {$i_4$};
	    \node[draw=none] (q1) at (3.5,0) {$i_5$};
	    \node[draw=none] (q2) at (3.5,-0.3) {$i_6$};
	    \node[rectangle,minimum width = 3em, minimum height = 3em, draw] (t4) at (1,0) {$\hat{U}$};
	    \node[rectangle,minimum width = 3em, minimum height = 3em, draw] (q4) at (2.5,0) {$\hat{V}^\dagger$};
	    \draw [thick]

	    (t0) -- (t0 -|  t4.west) 
	    (t1) -- (t1 -|  t4.west) 
	    (t2) -- (t2 -|  t4.west) 
	    (q0) -- (q0 -|  q4.east) 
	    (q1) -- (q1 -|  q4.east) 
	    (q2) -- (q2 -|  q4.east) 

	    (t4) -- (q4 -|  q4.west);

		\draw[line width=0.3mm, transform canvas={yshift=-1.5pt}] (t1) -- (u1);
		\draw[line width=0.3mm, transform canvas={yshift=1.5pt}] (u1) -- (t1);

	\end{tikzpicture}
	\caption{The decomposition of a tensor $T$ into a contraction of two tensors.}
	\label{fig:svdFlat}
\end{figure}

To apply the SVD to tensors with more than two indices we transform the tensor into a matrix and decompose that instead.
More specifically, for some tensor $T$ with indices $i_1, i_2, ..., i_N$ we produce two groups of indices $i_{j_1}, i_{j_2}, ..., i_{j_M}$ and $i_{j_M}, i_{j_{M+1}}, ..., i_{j_{N-M}}$ and identify
\begin{equation}
	M_{\{i_{j_1}, i_{j_2}, ..., i_{j_M}\}, \{i_{j_M}, i_{j_{M+1}}, ..., i_{j_{N-M}}\}} = T_{i_1, i_2, ..., i_N},
	\label{eq:flatten}
\end{equation}
where the notation $\{...\}$ means we merge (flatten) the listed indices into a single index.
This identification is depicted graphically in Fig.\ \ref{fig:flatten}.
The matrix $M$ may then be decomposed exactly according to equation\ \eqref{eq:svd} or approximately according to equation\ \eqref{eq:approxSVD}.
In either case after decomposition the original indices may be made separate once more, and the resulting diagram is as shown in Fig.\ \ref{fig:svdFlat}.

\subsection{Tree Decomposition}

\begin{figure}
	\begin{tikzpicture}
	    \node[draw=none] (t0) at (-2.5,0.3) {$i_1$};
	    \node[draw=none] (t1) at (-2.5,0) {$i_2$};

	    \node[draw=none] (t2) at (0,-0.3) {$i_3$};

	    \node[draw=none] (q0) at (3.5,0.3) {$i_4$};
	    \node[draw=none] (q1) at (3.5,0) {$i_5$};
	    \node[draw=none] (q2) at (3.5,-0.3) {$i_6$};

	    \node[rectangle,minimum width = 2em, minimum height = 2em, draw] (p4) at (-1,0.15) {$A$};
	    \node[rectangle,minimum width = 3em, minimum height = 3em, draw] (t4) at (1,0) {$B$};
	    \node[rectangle,minimum width = 3em, minimum height = 3em, draw] (q4) at (2.5,0) {$\hat{V}^\dagger$};
	    \draw [thick]

	    (t0) -- (t0 -|  p4.west) 
	    (t1) -- (t1 -|  p4.west) 
	    (p4.east) -- (p4.east -| t4.west)
	    (t2) -- (t2 -|  t4.west) 
	    (q0) -- (q0 -|  q4.east) 
	    (q1) -- (q1 -|  q4.east) 
	    (q2) -- (q2 -|  q4.east) 

	    (t4) -- (q4 -|  q4.west);

	\end{tikzpicture}
	\caption{The decomposition of a tensor into a contraction of three tensors.}
	\label{fig:secondSVD}
\end{figure}
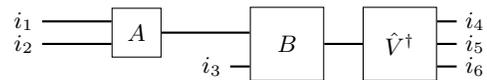

Tree decompositions are a natural extension of the singular value decomposition, in that they simply amount to an iterated SVD.
Using the example of Fig.\ \ref{fig:flatten}, a first application of the SVD produces the result in Fig.\ \ref{fig:svdFlat}.
Each of $\hat{U}$ and $\hat{V}^\dagger$ may then be decomposed further.
For instance we may group indices $i_1$ and $i_2$ together so that
\begin{equation}
	M_{\{i_1, i_2\}, \{i_3, i_i\}} = \hat{U}_{i_1, i_2, i_3, i_i},
\end{equation}
where $i_i$ is the internal line.
The matrix $M$ may then be decomposed with the SVD.
The result is shown in Fig.\ \ref{fig:secondSVD} where $A$ and $B$ result from the decomposition of $U$.

This process may be iterated until there are no tensors in the graph with more than three indices.
This choice was made because the cost of storing tensors scales exponentially in their rank and three is the minimal rank which allows a tree structure\footnote{In some cases this may incur unnecessary overhead. For instance if the indices of a tensor of rank four are maximally correlated attempting to decompose the tensor into two tensors of rank three doubles the storage required. However this cost is a constant overhead and so is typically acceptable.}
Furthermore the choice to proceed until all tensors have three indices is the only choice which can represent tensors of any rank greater than two without invoking tensors of varying ranks, and this uniform structure is considerably simpler to manipulate than a more general one with variable tensor ranks.

At this point further decomposition may be warranted on the original indices, but further decomposition on indices which are internally contracted just introduces further intermediate tensors with no corresponding reduction in rank.
This is because the SVD is already optimal, and so if an index results from performing the SVD it cannot be further compressed without raising the error threshold.
Because the SVD just introduces intermediate tensors between sets of indices it cannot introduce cycles into the graph and so the result of this iterative decomposition is a tree.
What we mean by a tensor tree then is an acyclic tensor network (graph) composed of only tensors with three or fewer indices.

\subsection{Choice Sensitivity}

\begin{figure}
	\begin{tikzpicture}
	    \node[draw=none] (v0) at (-3,3.3) {$i$};
	    \node[draw=none] (v2) at (-3,2.7) {$j$};
	    \node[draw=none] (u0) at (0.5,3.3) {$k$};
	    \node[draw=none] (u2) at (0.5,2.7) {$l$};
	    \node[rectangle,minimum width = 3em, minimum height = 3em, draw] (v4) at (-2,3) {$A$};
	    \node[rectangle,minimum width = 3em, minimum height = 3em, draw] (u4) at (-0.5,3) {$B$};
	    \draw [thick]
	    (v0) -- (v0 -|  v4.west) 
	    (v2) -- (v2 -|  v4.west) 
	    (v4.east) -- (v4.east -| u4.west)
	    (u0) -- (u0 -|  u4.east) 
	    (u2) -- (u2 -|  u4.east); 

	    \node[draw=none] (vv0) at (-3,1.8) {$i$};
	    \node[draw=none] (vv2) at (-3,1.2) {$k$};
	    \node[draw=none] (uu0) at (0.5,1.8) {$j$};
	    \node[draw=none] (uu2) at (0.5,1.2) {$l$};
	    \node[rectangle,minimum width = 3em, minimum height = 3em, draw] (vv4) at (-2,1.5) {$A'$};
	    \node[rectangle,minimum width = 3em, minimum height = 3em, draw] (uu4) at (-0.5,1.5) {$B'$};
	    \draw [thick]
	    (vv0) -- (vv0 -|  vv4.west) 
	    (vv2) -- (vv2 -|  vv4.west) 
	    (vv4.east) -- (vv4.east -| uu4.west)
	    (uu0) -- (uu0 -|  uu4.east) 
	    (uu2) -- (uu2 -|  uu4.east); 

	    \node[draw=none] (vvv0) at (-3,0.3) {$i$};
	    \node[draw=none] (vvv2) at (-3,-0.3) {$l$};
	    \node[draw=none] (uuu0) at (0.5,0.3) {$k$};
	    \node[draw=none] (uuu2) at (0.5,-0.3) {$j$};
	    \node[rectangle,minimum width = 3em, minimum height = 3em, draw] (vvv4) at (-2,0) {$A''$};
	    \node[rectangle,minimum width = 3em, minimum height = 3em, draw] (uuu4) at (-0.5,0) {$B''$};
	    \draw [thick]
	    (vvv0) -- (vvv0 -|  vvv4.west) 
	    (vvv2) -- (vvv2 -|  vvv4.west) 
	    (vvv4.east) -- (vvv4.east -| uuu4.west)
	    (uuu0) -- (uuu0 -|  uuu4.east) 
	    (uuu2) -- (uuu2 -|  uuu4.east); 

	\end{tikzpicture}
	\caption{The three decompositions of a tensor with four indices.}
	\label{fig:svd4}
\end{figure}
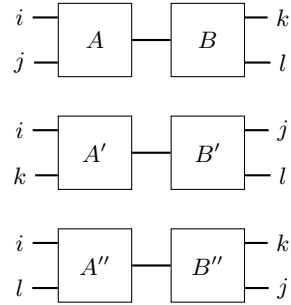

\begin{figure}
	\includegraphics[width=0.45\textwidth]{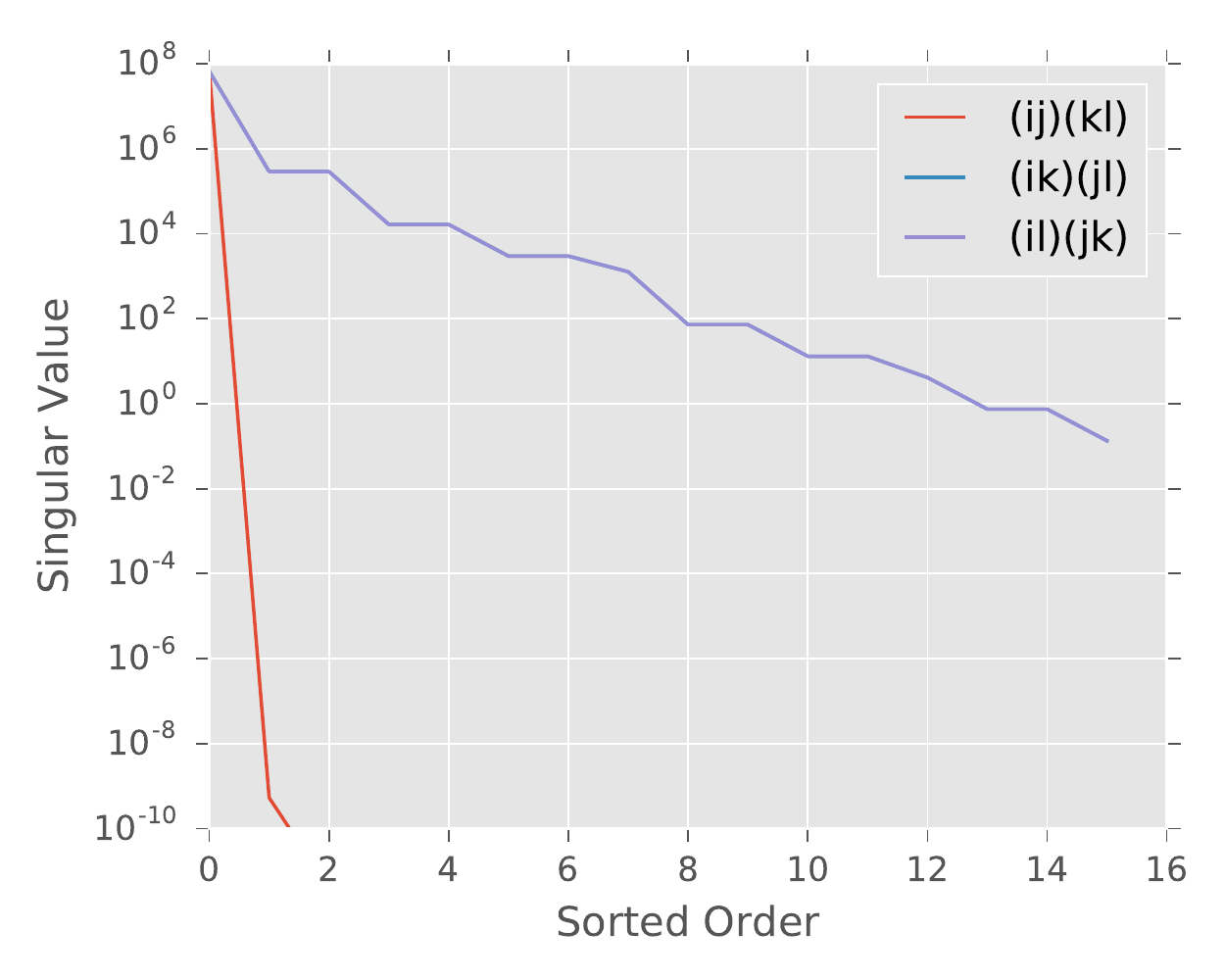}
	\caption{The singular values of three decompositions of a tensor are plotted against their sorted order. Note that two decompositions are formally the same for this example and so yield the same spectrum of singular values.}
	\label{fig:ex1}
\end{figure}

To understand why the tree decomposition is sensitive to the precise choice of tree, consider the tensor
\begin{equation}
	T_{ijkl} = e^{ij}e^{kl}.
	\label{eq:ex1}
\end{equation}
There are three possible tree decompositions of $T$, shown in Fig.\ \ref{fig:svd4}.
The corresponding singular values for each are plotted in Fig.\ \ref{fig:ex1}.
Because $T$ factors precisely into a product of terms dependent on $i$ and $j$ and ones dependent on $k$ and $l$, the first decomposition has only one non-zero singular value.
In the numerical decomposition the second largest singular value is a factor of $10^{16}$ smaller than the largest, which is within the floating point tolerance.
By contrast the other two decompositions have many non-zero singular values, and may require several of these to be included to produce a good approximation.
So for instance an approximation which is good to one part in $10^{4}$ requires five singular values to be retained with these decompositions.

Picking a good decomposition only becomes more important for tensors with more indices.
This is because each internal index must encode the correlations between all indices on either side of it.
Separating a highly correlated pair of indices therefore requires increasing the rank of every intermediate index between them to accommodate their correlations.

\section{Algorithms}

\subsection{Brute Force}
\label{sec:brute}

A straightforward way to optimize over a finite set of choices is to evaluate every choice and pick the best one.
For tensors with only a few indices this is possible: we simply enumerate all possible tree decompositions, evaluate each one, and pick the one which requires the storing the fewest elements for a given error threshold $\epsilon$.
This enumeration may be done quite efficiently\ \citep{Yamanaka2009, doi:10.1137/0208006}, but unfortunately the number of trees grows rapidly with the number of indices.
The number of trees with $N$ leaves and three links on each internal node is
\begin{equation}
	S_N = (2n-5)!! = 2^{2-N}\frac{(2N-4)!}{(N-2)!}
\end{equation}
\citep{doi:10.1093/tropej/fmn004}.
For large $N$,
\begin{equation}
S_{N} \sim 2^{-N}\frac{(2N)!}{N!}.
\end{equation}
Applying Stirling's approximation we find
\begin{equation}
S_{2N-2,N} \sim 2^{-N} \left(\frac{2N}{e}\right)^{2N}\left(\frac{N}{e}\right)^{-N} = \left(\frac{2N}{e}\right)^{N}
\end{equation}
which grows exponentially with $N$.
As such this method is only practical for tensors with a small number of indices.

Our implementation of this algorithm iterates over all trees.
It identifies and performs a sequence of SVD operations to produce each tree.
When the truncation error is negligible the order of SVD operations does not matter.
In testing however there were rare cases in which the order of application of the SVD mattered.
This may occur, for instance, when a singular value lies very near to the threshold and so can be pushed to one side or the other by truncation error.
Still, in almost all cases the brute force algorithm identifies the optimal tree decomposition, and so we do not further examine all possible SVD orderings.
As it stands the algorithm is practical only for relatively small tensors, and examining SVD orderings would only make this worse.

\subsection{Greedy Choice}
\label{sec:algo}

While considering every possible tree decomposition ensures that we find the best one, there are useful heuristics which can often do nearly as well.
One such heuristic is to construct the tree by locally minimizing the number of terms left in the truncated SVD at each stage.
More specifically, we consider all factorizations of the tensor $T$ with indices $i_1,i_2,...,i_N$ of the form
\begin{equation}
	T_{i_1,i_2,...,i_N} \approx A_{i_{j_1},i_{j_2},i_i} T'_{i_i,i_{j_3},...,i_{j_N}},
	\label{eq:decomp}
\end{equation}
where there is summation over the repeated internal index $i_i$.
Thus we are looking for the pair of indices to factor away from the rest which minimize the rank of the resulting internal index.
This requires checking $N(N-1)/2$ factorizations except in the special case where $N = 4$, where symmetry allows us to check just $3$ cases.
Once we have found the best pair to factor away from $T$ we may iterate and search for the best index pair to factor out from $T'$.
This process ends when the process returns two tensors each with three indices.

It is possible to further improve the performance of this algorithm by noting that most indices of $T'$ are the same as those of $T$.
If $T$ is exactly represented by the contraction of $A$ and $T'$ then we do not need to check pairs of these again, as they represent precisely the same decomposition as before.
If $T$ is only approximately represented then in principle the rank of these pairs may have changed, but in practice for a sufficiently small error threshold $\epsilon$ we expect this to be unlikely.
As a result we only need to examine pairs involving the new index $i_i$, of which there are $N-2$.
Iterating until there are just three indices on each tensor requires
\begin{equation}
\frac{1}{2}N(N-1) + \sum_{j=1}^{N-4} (j+2) = N^2 - 2N - 2
\end{equation}
pairs to be tested if $N > 4$.
In the limit of large $N$ this scales quadratically, and so is immensely preferable to the exponential scaling of the brute force algorithm.

Note that for simplicity our implementation of this algorithm does not include this improvement, and has cubic runtime in $N$.
Regardless the number of comparisons is polynomial in $N$ rather than exponential, which is the key advantage of this algorithm.

It is important to emphasize that this algorithm does not generally produce the optimal tree decomposition, though in specific instances it may.
For instance if the tensor was formed from an outer product of rank-2 objects this algorithm would correctly identify the pairs of indices which are uncorrelated from all others and the resulting tensor tree would be composed of the original rank-2 objects, augmented by links of dimension one to one another and possibly rescaled.
Likewise if the tensor was formed by contracting a sequence of identical rank-3 tensors against one another, as in the $N$-point correlation function of the one-dimensional Ising model (see Section~\ref{sec:numerics}), the entanglement between any pair of indices and the remainder of the indices is guaranteed to be monotonic in the distance within the pair and hence a greedy approach correctly identifies that adjacent indices ought to be split off from the rest.
These scenarios are quite special, however, and do not necessarily generalize.
Hence the greedy algorithm should be treated as producing approximations to the best choice of decomposition rather than the best decomposition outright.

\subsection{Runtime Comparison}

In both algorithms the runtime of the SVD typically scales exponentially in $N$.
This is because the number of entries in the matrix scales as $d^N$, where $d$ is the typical dimension of the indices.
This may be avoided by employing approximate decompositions and rank estimations~\citep{Ballani2014}.
Nevertheless such approximations are not our focus and so we proceed with the naive exponential implementation.

Because the runtime is proportional to the cost of performing decompositions and/or rank estimations, for comparing these two algorithms we may just examine how many such operations must be performed.
In the brute-force algorithm this scales as $(2N/e)^N$, while in the greedy algorithm it scales as $N^2$.
For $N > 1$ the latter is smaller, and even for modest $N$ the difference in runtime may be enormous.
For example when $N=10$ the brute force algorithm requires approximately $(20/e)^{10}\approx 5\times 10^8$ comparisons while the greedy one just requires of order $10^2$ or $10^3$ comparisons depending on the implementation.
This makes the latter preferable so long as it produces results of comparable quality.

\section{Numerical Experiments}
\label{sec:numerics}

\begin{figure}
	\includegraphics[width=0.45\textwidth]{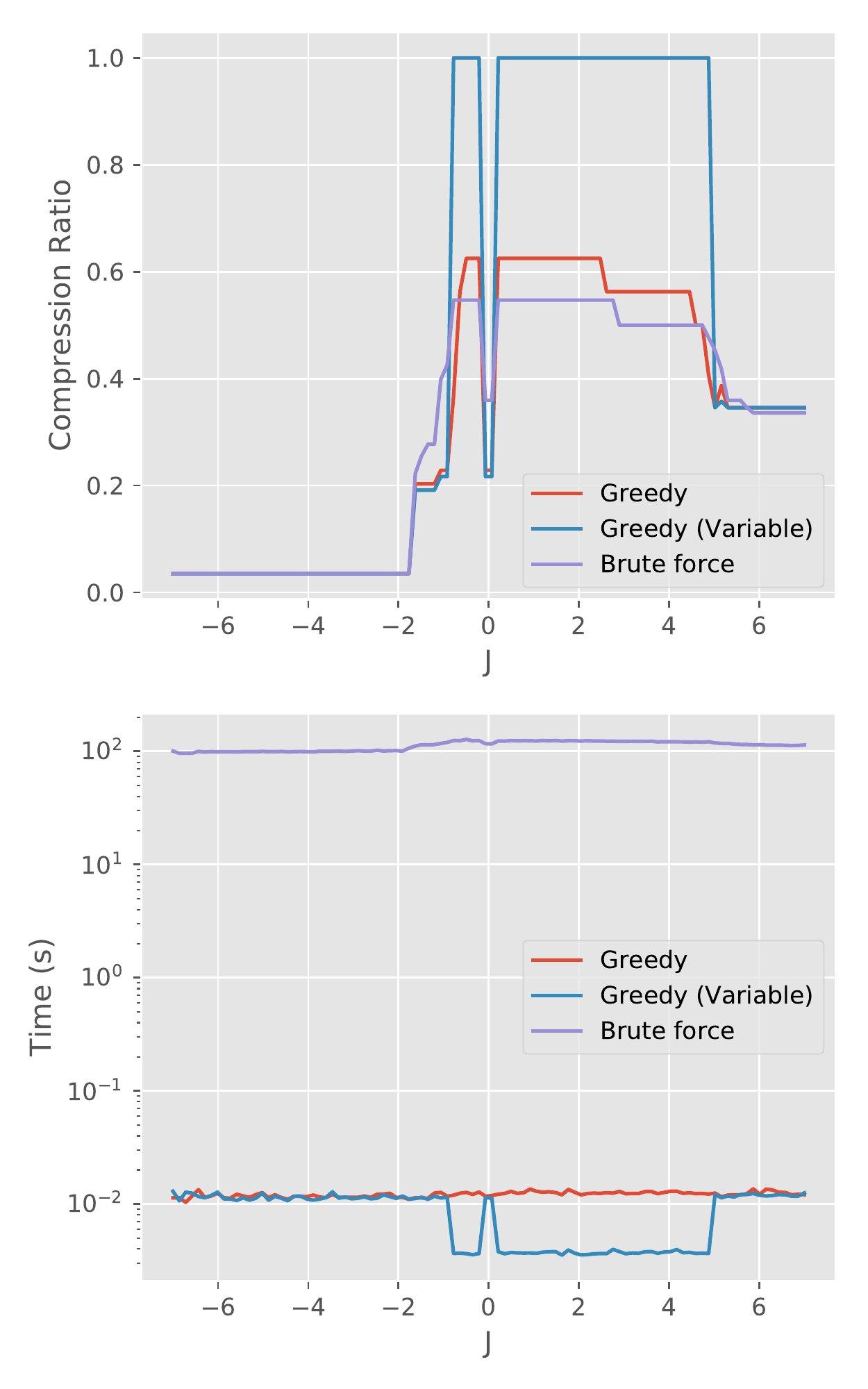}
	\caption{The compression ratio (top), defined as the number of values stored in the tree decomposition divided by the number of entries in the tensor, is shown as a function of the nearest neighbour Ising coupling $J$ for both the greedy (blue) and brute force (red) algorithms. On the bottom the time required by each algorithm is shown. The tensors used were 2D Ising $N$-point functions for a system formed of a $3\times 3$ square lattice with open boundary conditions. Note that the two algorithms produce very similar results, and that the compression ratio is worst for intermediate $J$. Furthermore note that this example runs into one of the edge cases where the order of the SVD matters, as can be seen by the greedy algorithm outperforming the brute force algorithm for two values of $J$.}
	\label{fig:ex6}
\end{figure}

We have already discussed the relative runtime of our algorithms and so we proceed directly to examining their relative performance on a variety of tasks.
For simplicity our implementation of the greedy algorithm does not take advantage of the fact that index pairs repeat between iterations.
Furthermore we take the output from the brute force algorithm to be the standard of comparison for the greedy algorithm, as it is optimal in all but certain rare cases~\footnote{Because the brute force algorithm relies on the iterated application of SVD it may fail to produce optimal results if approximation error from an early SVD application suffices to change the effective ranks of subsequent applications. In such an instance this error could result in a sub-optimal tree being selected. With $\epsilon$ set as small as in our numerical experiments this ought to be rare, and indeed we only see it occurring twice. Both instances appear as the greedy algorithm outperforming the brute force one in Figure~\ref{fig:ex6}.}.
Timing was done on an Intel IvyBridge processor using the Intel MKL linear algebra backend.
The linear algebra was parallelized across cores when this provided a speed improvement (i.e. for large tensors) but the search algorithms were run serially.
This prevents competition between threads for linear algebra resources.

For the threshold we use
\begin{equation}
	\epsilon^2 = 10^{-6} \sum_i \Sigma_{ii}^2.
\end{equation}
This ensures that the relative $L_2$ error incurred with each eliminated singular value is no more than $10^{-6}$, and means that the normalization of the tensor does not impact our choice of decomposition.

To begin we consider the example of equation\ \eqref{eq:ex1}.
We tested the case where all indices range from $0$ to $3$ inclusive and found that the greedy algorithm identified the optimal decomposition.
This is not surprising, as the greedy algorithm reduces to the brute force algorithm when there are just four indices.

\begin{figure}
	\begin{tikzpicture}
	    \node[draw=none] (t0) at (-2.5,0.3) {$i$};
	    \node[draw=none] (t1) at (-2.5,-0.3) {$j$};
	    \node[draw=none] (t2) at (0.7,1.3) {$m$};

	    \node[draw=none] (q0) at (3.8,0.3) {$k$};
	    \node[draw=none] (q2) at (3.8,-0.3) {$l$};

	    \node[rectangle,minimum width = 3em, minimum height = 3em, draw] (p4) at (-1,0) {$A$};
	    \node[rectangle,minimum width = 3em, minimum height = 3em, draw] (t4) at (0.7,0) {$B$};
	    \node[rectangle,minimum width = 3em, minimum height = 3em, draw] (q4) at (2.5,0) {$C$};
	    \draw [thick]

	    (t0) -- (t0 -|  p4.west) 
	    (t1) -- (t1 -|  p4.west) 
	    (p4.east) -- (p4.east -| t4.west)
	    (t2) -- (t2 |-  t4.north) 
	    (q0) -- (q0 -|  q4.east) 
	    (q2) -- (q2 -|  q4.east) 

	    (t4) -- (q4 -|  q4.west);

	\end{tikzpicture}
	\caption{The decomposition of a tensor into a contraction of three tensors.}
	\label{fig:ex2diag}
\end{figure}
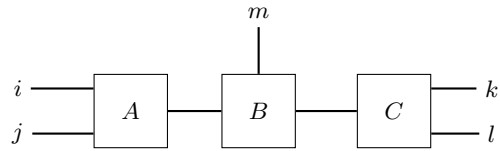

We further considered the case
\begin{equation}
	T_{ijklm} = e^{ij}e^{kl}e^m,
	\label{eq:T}
\end{equation}
where once more all indices range from $0$ to $3$ inclusive.
Once more the greedy algorithm reproduces the result from the brute force algorithm, which is shown in Figure\ \ref{fig:ex2diag}.
Indeed the links between $A$ and $B$ and between $B$ and $C$ are one-dimensional in this example, reflecting the total factorization of equation~\eqref{eq:T}.
This results in an overall compression from $4^5=1024$ entries to just $36$, and is an example of the best case for this sort of tensor compression.

The decomposition in Figure\ \ref{fig:ex2diag} remains optimal even for more complex tensors like
\begin{equation}
	T_{ijklm} = e^{ij}e^{kl}e^m + \sin(ij)\cos(kl)\tanh(m)
\end{equation}
because the required bond dimension to encode the correlations between $m$ and the remaining indices is less than that required to encode those between either of $i$ and $j$ or $k$ and $l$.
The greedy algorithm once more performs optimally in this case.

\begin{figure}
	\includegraphics[width=0.45\textwidth]{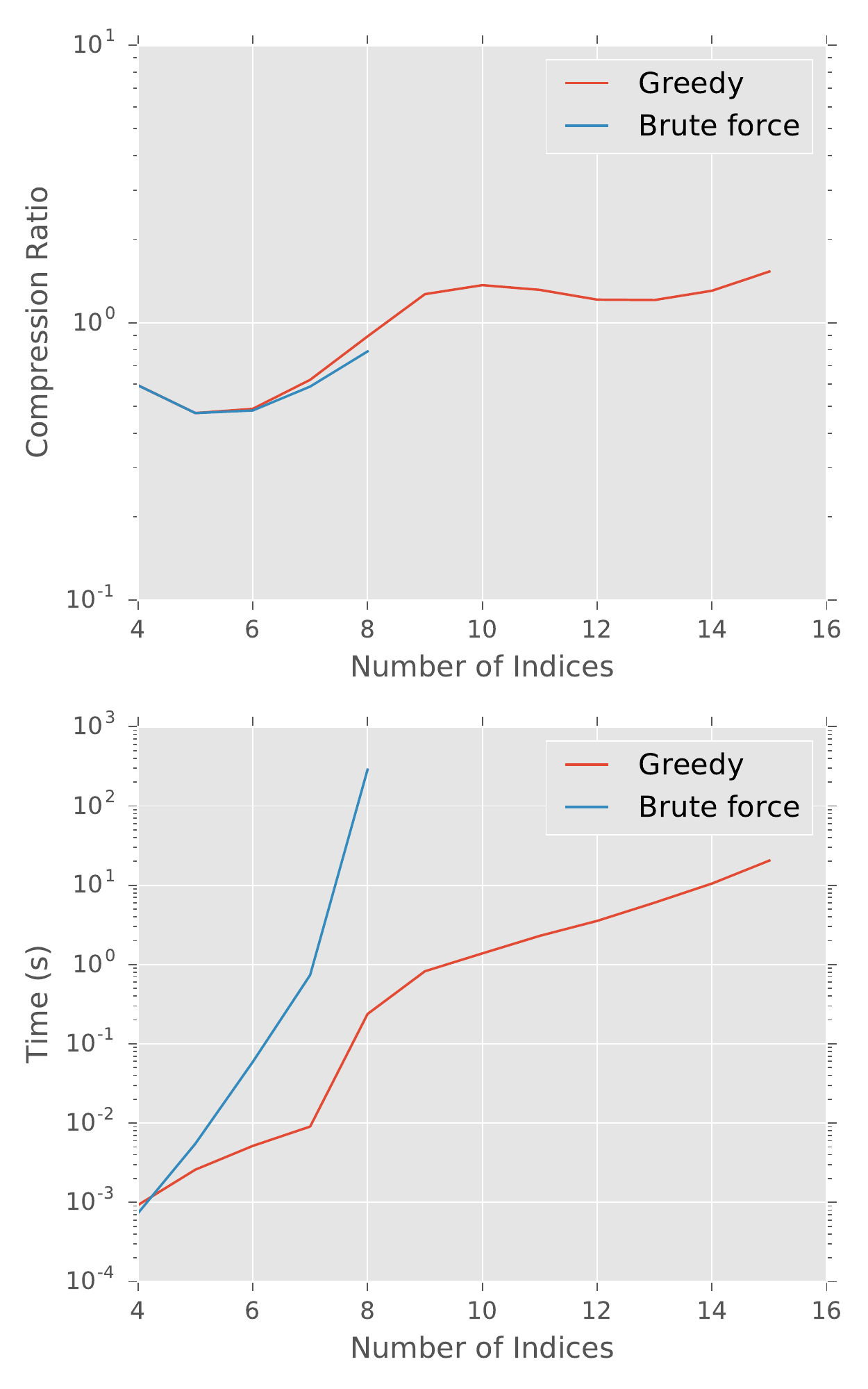}
	\caption{The compression ratio (top), defined as the number of values stored in the tree decomposition divided by the number of entries in the tensor, is shown as a function of the number of indices $N$ of the tensor for both the greedy (blue) and brute force (red) algorithms. On the bottom the time required by each algorithm is shown. The tensors were randomly filled with $0$ and $1$ with probabilities $1-p$ and $p$ respectively and tensors with no non-zero entries were rejected from the sample. In this case $p=0.05$. The results were averaged over $100$ valid tensors at each $N$. Note that the greedy algorithm is shown over a wider range of $N$ because the brute force algorithm is too slow to be used at larger $N$.}
	\label{fig:ex2}
\end{figure}

To construct more challenging examples we next considered the family of tensors with $N$ indices of dimension $2$ where each entry is randomly and independently chosen as either $1$ with probability $p$ or $0$ with probability $1-p$.
In the low-$p$ limit tensors drawn from this distribution exhibit non-uniform correlations across all indices because there are few correlations overall owing to most entries vanishing, and so clusters and hence correlations are typical.
Figure\ \ref{fig:ex2} shows the factor by which each algorithm was able to compress tensors generated in this fashion as a function of the number of indices $N$ as well as the time required to find this decomposition.
For the cases where both algorithms have tractable runtimes they perform very similarly, with the greedy algorithm usually finding optimal or near-optimal solutions.
For larger $N$ the brute force algorithm becomes intractable but the greedy algorithm remains performant.
This can be seen in the timing curves, where the difference begins at several orders of magnitude and rises rapidly in $N$ even on a logarithmic scale.
Beyond $N=8$ it was not practical to evaluate this algorithm with enough samples for robust averaging, and so the brute force benchmark ends there.

One feature of Figure\ \ref{fig:ex2} worth noting is that this class of tensors becomes less compressible as $N$ increases, and for some $N$ the decompositions are actually less efficient than just directly storing the original tensor.
This is not just a feature of the greedy algorithm, as the optimal result shows the same trend as far as it has been evaluated.
To understand this note that for a given pair of indices the tensor may be viewed as a collection of matrices, each of which possesses those two indices.
In the large-$N$ limit the number of such matrices increases rapidly.
This increase in the mutual entropy between any pair of indices increases the SVD rank required to maintain the specified accuracy.
When this rank approaches the original matrix rank the SVD can actually increase the number of entries being stored from $nm$ up to $(n+m)\min(n,m)$, where $n$ and $m$ are the matrix dimensions.
This results in an overall increase in the memory required to store the tensor.

\begin{figure}
	\includegraphics[width=0.45\textwidth]{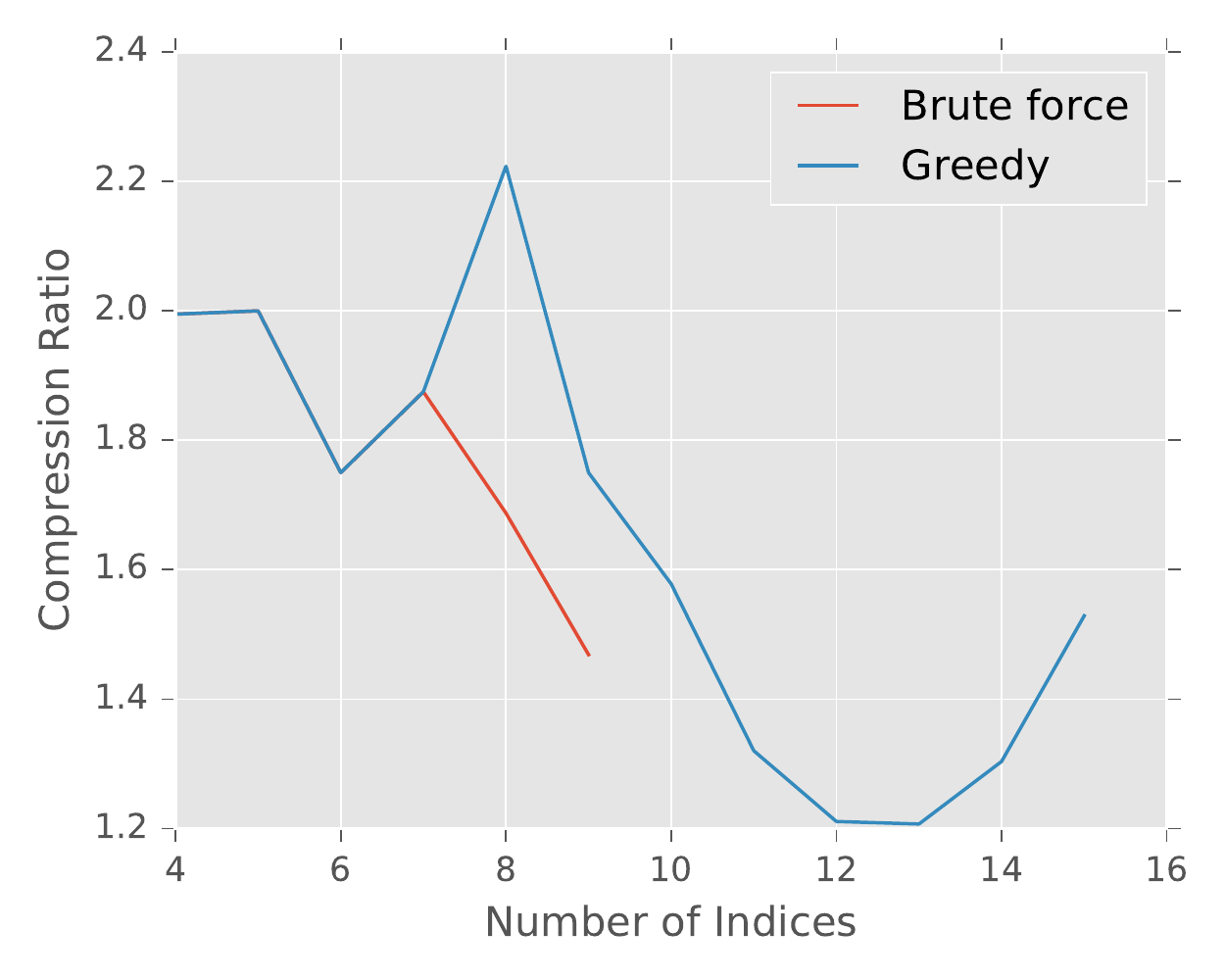}
	\caption{The compression ratio, defined as the number of values stored in the tree decomposition divided by the number of entries in the tensor, is shown as a function of the number of indices $N$ of the tensor for both the greedy (blue) and brute force (red) algorithms. The tensors were randomly filled with values drawn independently from a random normal distribution with mean $0$ and variance $1$. The results were averaged over $100$ tensors at each $N$. Note that the greedy algorithm is shown over a wider range of $N$ because the brute force algorithm is too slow to be used at larger $N$.}
	\label{fig:ex3}
\end{figure}

This incompressibility is actually a common feature of random tensors.
For instance Figure\ \ref{fig:ex3} shows the same plot as in the top panel of Figure\ \ref{fig:ex2} but for tensors whose entries are drawn randomly and independently from a normal distribution.
For all $N$ the compression ratio is quite poor, in stark contrast to the more structured examples.
The two algorithms perform identically for $N < 7$, and for $N=8,9$ the brute force algorithm is somewhat better.
At larger $N$ the compression ratio of the greedy algorithm improves and plateaus in performance near a compression ratio of unity.
We expect that unbiased random tensors ought not to be compressible so this result is likely close to optimal.

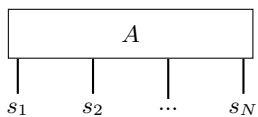
\begin{figure}
	\begin{tikzpicture}
	    \node[draw=none] (t0) at (0,-1) {$s_1$};
	    \node[draw=none] (t1) at (1,-1) {$s_2$};
	    \node[draw=none] (t2) at (2,-1) {$...$};
	    \node[draw=none] (t3) at (3,-1) {$s_N$};

	    \node[rectangle,minimum width = 10em, minimum height = 2em, draw] (p0) at (1.5,0) {$A$};
	    \draw [thick]

	    (t0) -- (t0 |- p0.south)
	    (t1) -- (t1 |- p0.south)
	    (t2) -- (t2 |- p0.south)
	    (t3) -- (t3 |- p0.south);

	\end{tikzpicture}
	\caption{The Ising $N$-point probability function depicted as a tensor.}
	\label{fig:ising}
\end{figure}

A case with more physical content is that of a tensor encoding the $N$-point correlation function for a lattice system.
For instance in the 1D classical Ising model with open ends states are weighted as
\begin{align}
	P(\{s_i\}) \propto e^{-J \sum_{i=0}^{N-1} s_i s_{i+1}},
\end{align}
where each $s_i$ is drawn from $\{0,1\}$.
Interpreting each $s_i$ as an index produces a tensor with $N$ indices, shown in Figure\ \ref{fig:ising}.
This object ought to be quite compressible because the correlations it encodes are local and so factor more readily.

\begin{figure}
	\begin{tikzpicture}
	    \node[draw=none] (t0) at (0,-1) {$s_1$};
	    \node[draw=none] (t1) at (1,-1) {$s_2$};
	    \node[draw=none] (t2) at (2,-1) {$s_3$};
	    \node[draw=none] (t3) at (3,-1) {$s_4$};
	    \node[draw=none] (t4) at (4,-1) {$s_5$};
	    \node[draw=none] (t5) at (5,-1) {$s_6$};
	    \node[draw=none] (t6) at (6,-1) {$s_7$};
	    \node[draw=none] (t7) at (7,-1) {$s_8$};
	    \node[draw=none] (t8) at (8,-1) {$s_9$};

	    \node[rectangle,minimum width = 4em, minimum height = 2em, draw] (p0) at (7.5,0) {$T_{89}$};
		\draw [thick]
			(t7) -- (t7 |- p0.south)	
			(t8) -- (t8 |- p0.south);	

	    \node[rectangle,minimum width = 6em, minimum height = 2em, draw] (p1) at (6.7,1) {$T_{789}$};
		\draw [thick]
			(t6) -- (t6 |- p1.south)	
			(p0) -- (p0 |- p1.south);	

	    \node[rectangle,minimum width = 7em, minimum height = 2em, draw] (p2) at (6,2) {$T_{6789}$};
		\draw [thick]
			(t5) -- (t5 |- p2.south)	
			(p1) -- (p1 |- p2.south);	

	    \node[rectangle,minimum width = 7em, minimum height = 2em, draw] (p3) at (5,3) {$T_{56789}$};
		\draw [thick]
			(t4) -- (t4 |- p3.south)	
			(p2) -- (p2 |- p3.south);	

	    \node[rectangle,minimum width = 7em, minimum height = 2em, draw] (p4) at (4,4) {$T_{456789}$};
		\draw [thick]
			(t3) -- (t3 |- p4.south)	
			(p3) -- (p3 |- p4.south);

	    \node[rectangle,minimum width = 7em, minimum height = 2em, draw] (p5) at (3,5) {$T_{3456789}$};
		\draw [thick]
			(t2) -- (t2 |- p5.south)	
			(p4) -- (p4 |- p5.south);

	    \node[rectangle,minimum width = 9em, minimum height = 2em, draw] (p6) at (1.5,6) {$T_{123456789}$};
		\draw [thick]
			(t0) -- (t0 |- p6.south)
			(t1) -- (t1 |- p6.south)	
			(p5) -- (p5 |- p6.south);

	\end{tikzpicture}
	\caption{The Ising $N$-point probability function decomposed into a tensor tree for $N=9$ and $J=1$. Note that the structure is regular and local, with indices which are neighbours in the Ising model appearing near one another in the tensor tree. This reflects the local nature of correlations in the Ising model.}
	\label{fig:isingTree}
\end{figure}
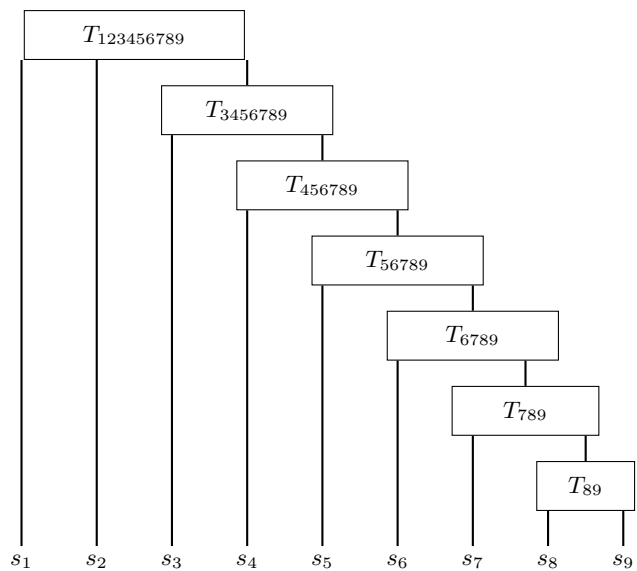

Figure\ \ref{fig:isingTree} shows the tree decomposition produced by the brute force algorithm for the Ising $N$-point function with $N=9$ and $J=1$.
Note that the structure is regular and local, with indices which are neighbours in the Ising model appearing near one another in the tensor tree.
This reflects the local nature of correlations in the Ising model.

\begin{figure}
	\includegraphics[width=0.45\textwidth]{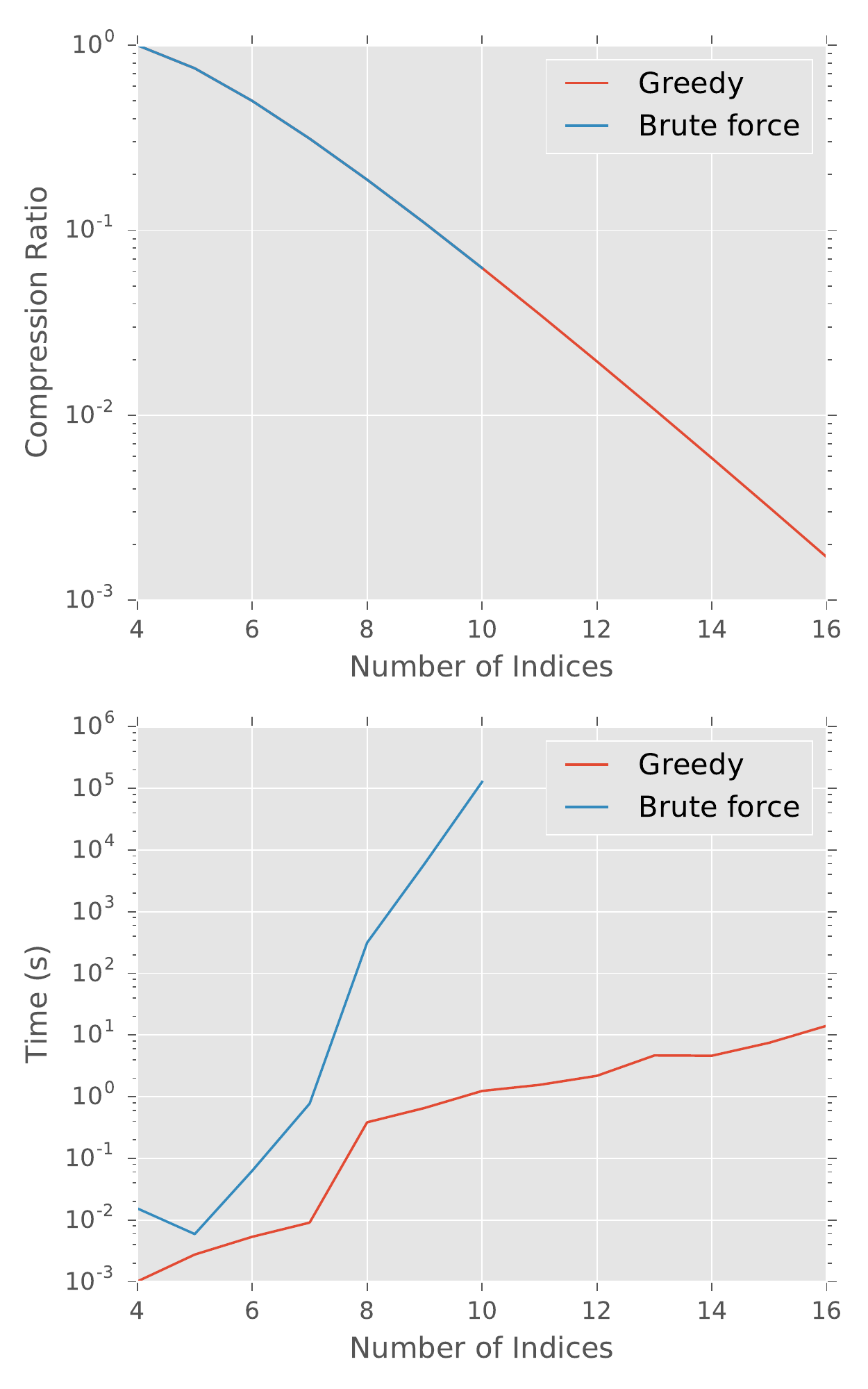}
	\caption{The compression ratio (top), defined as the number of values stored in the tree decomposition divided by the number of entries in the tensor, is shown as a function of the number of indices $N$ of the tensor for both the greedy (blue) and brute force (red) algorithms. On the bottom the time required by each algorithm is shown. The tensors used were Ising $N$-point functions with $J=1$. Note that the two algorithms produce identical results.}
	\label{fig:ex4}
\end{figure}

Figure\ \ref{fig:ex4} shows the compression ratio of the Ising $N$-point function as a function of system size.
The two algorithms produce identical results, so in these cases the greedy algorithm finds the optimal tree decomposition.
Once more the brute force algorithm proved impossible to benchmark beyond moderate $N$ and so only the greedy algorithm is shown for larger values.

Interestingly, the compression ratio in Figure\ \ref{fig:ex4} decreases rapidly as a function of $N$, in keeping with the intuition that Ising correlations are highly redundant.
In particular, the decrease is exponential in $N$.
To understand this note that the tensor size is $2^N$.
For finite $J$ there is a corresponding correlation length $\xi(J)$.
The size of the compressed tensor is set by the dimensions of the internal bonds in the tree.
This in turn is set by the number of degrees of freedom above the given error threshold, which is proportional to the entanglement entropy between the two sides of the bond, and so scales as $2^{\xi(J)}$\footnote{For a more detailed version of this argument see e.g.~\citet{MolinaVilaplana2014}.}
As such the compression ratio scales like $2^{\xi(J) - N}$, which is consistent with the data in Figure\ \ref{fig:ex4}.

\begin{figure}
	\includegraphics[width=0.45\textwidth]{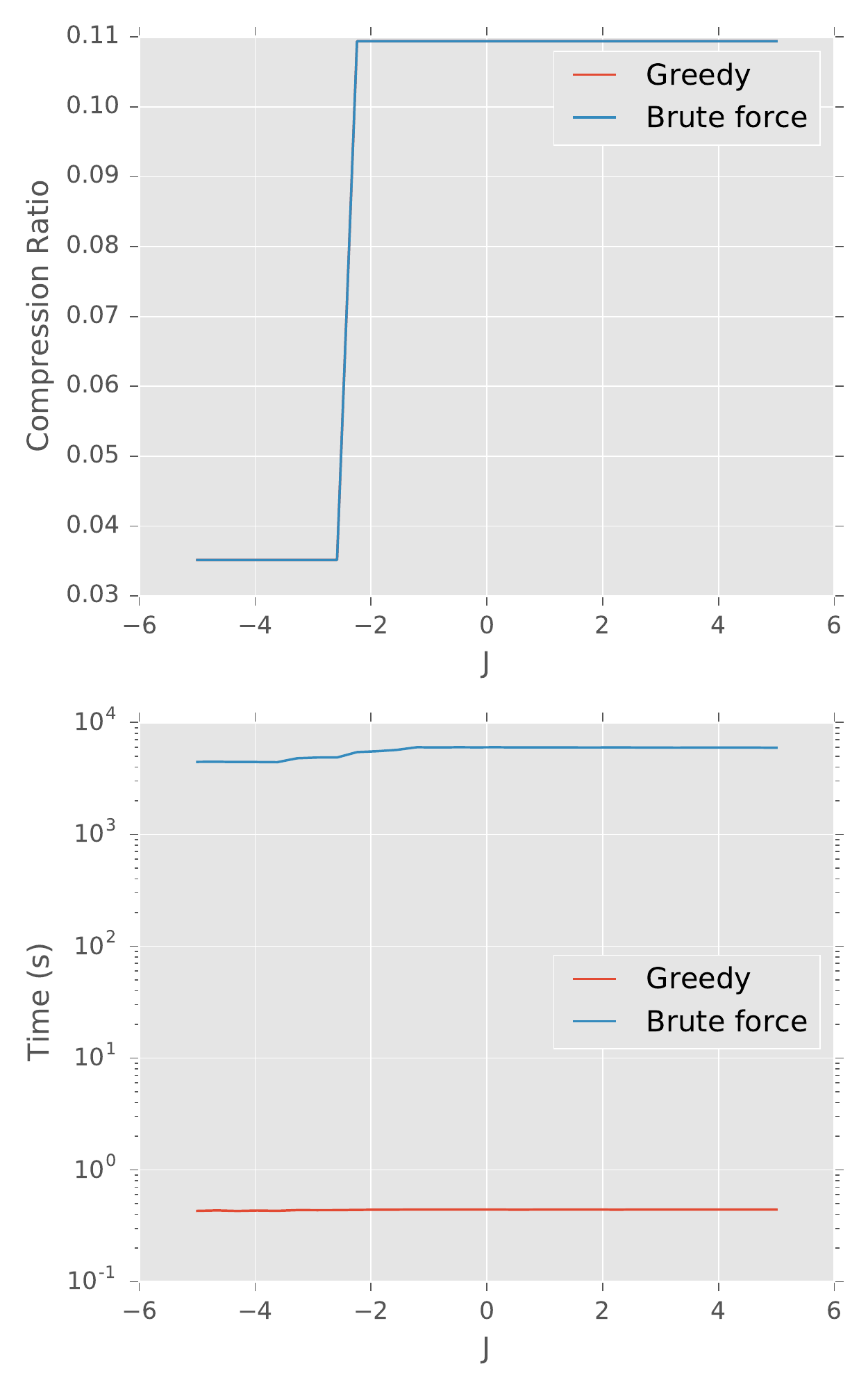}
	\caption{The compression ratio (top), defined as the number of values stored in the tree decomposition divided by the number of entries in the tensor, is shown as a function of the nearest neighbour Ising coupling $J$ for both the greedy (blue) and brute force (red) algorithms. On the bottom the time required by each algorithm is shown. The tensors used were Ising $N$-point functions for a system with $N=8$ sites. Note that the two algorithms produce identical results.}
	\label{fig:ex5}
\end{figure}

Figure\ \ref{fig:ex5} shows the same compression ratio and timing data as a function of the site-site coupling $J$.
The timing is largely independent of $J$, which indicates that it is mainly being set by the combinatorics of the problem and not by the underlying matrix transformations.
This highlights the need for efficient means of choosing a decomposition.

Once more the two algorithms produce identical results for the compression ratio, with variation only near the transition in compression ratio at $J=-2$.
The increase in compression ratio with $J$ occurs because as $J$ increases correlations become longer-ranged, and so become more difficult to encode in a local model such as a tree decomposition.
This manifests as a sharp transition because near $J=-2$ a specific singular value crosses the threshold $\epsilon$ and becomes relevant.

As a further test consider the 2D Ising probability function
\begin{align}
	P(\{s_i\}) \propto e^{-J \sum_{\langle ij \rangle} s_i s_j},
	\label{eq:P}
\end{align}
where $\langle ij \rangle$ indicates all nearest-neighbour pairs.
Figure\ \ref{fig:ex6} shows the compression ratio and timing for this function, again interpreted as a tensor indexed by $\{s_i\}$, as a function of $J$ for the case of a $3\times 3$ open-boundary square lattice.
In addition to the brute force and greedy algorithms a variant of the greedy algorithm which uses variable-size tensors was included.
This variant leaves tensors of rank greater than three in the final decomposition if the next step of their decomposition would not produce net compression.
In that sense it is also greedy in the space of possible final tensor ranks.

Once more the algorithms produce very similar results.
In this case though the compression ratio is worst for intermediate $J$.
This is matched with an increase in computation time, which makes sense because the SVD in this region is more time consuming.
At extreme $J$ the complexity falls because as the $J\rightarrow \pm \infty$ the ferromagnetic and antiferromagnetic ground states come to dominate the probability distribution.
This makes the tensors more compressible.

For most $J$ the regular greedy algorithm outperformed the variable-size one.
This reflects the fact that the latter may determine that the next decomposition step produces no net improvement, or indeed produces a worse compression ratio, than doing nothing at all.
For random tensors, or indeed for the tensor in equation~\eqref{eq:P} near $J=-1$ and $J=5$, this is a correct determination, but in other cases it can leads to premature termination of the decomposition and worse overall performance.

It is worth noting that in this case both greedy algorithms actually outperform the brute force algorithm in small regions near $J=-1$, $J=0$ and $J=7$.
These are among the cases where the order of the SVD matters because of the truncation error.

A key takeaway from these examples is that the structure of the tensor matters.
For instance it is not surprising that models with local structure such as these should be good use cases for the greedy algorithm because it is designed to take advantage of the existence a hierarchy of correlations, identifying the most tightly-correlated indices and moving outward from there.

Finally, for comparison purposes consider Example 6.4 of~\citet{Ballani2014}.
Let
\begin{align}
	g(x,y) \equiv \frac{1}{|y-\alpha x| + 1}
\end{align}
for fixed $\alpha$.
Furthermore given $N\in \mathbb{N}$ let
\begin{align}
	\xi_{i_1,...,i_N} \equiv \sum_{\mu=1}^N 2^{\mu-N} i_\mu - 1,
\end{align}
where each $i_k \in \{0,1\}$.
With this they define the tensor
\begin{align}
	A_{i_1,...,i_{2N}} \equiv g(\xi_{i_1,...,i_N}, \xi_{i_{N+1},...,i_{2N}}).
	\label{eq:A}
\end{align}
This example is non-trivial because neither the structure of $g$ nor the bitwise indexing of $\xi$ permit any obvious factoring.

Table~\ref{tab:compare} shows the effective dimension defined in~\citet{Ballani2014} for both their adaptive agglomeration scheme and the greedy one introduced in this work.
To match the parameters used in their tests the dimension was set to $N=8$, the tolerance $\epsilon = 10^{-8}$, and the test was performed for each $\alpha\in\{0,0.25,0.5,0.75,1.0\}$.
Furthermore to account for possible differences in how the methods determine the SVD tolerance the greedy algorithm was also run with a tolerance of $10^{-9}$.

In three cases, namely $\alpha = 0$, $\alpha=0.25$ and $\alpha=1.0$ the greedy algorithm outperformed the agglomeration algorithm.
This was true both for $\epsilon=10^{-8}$ and $\epsilon=10^{-9}$.
In the remaining cases the agglomeration algorithm outperformed the greedy one.
In no case was the performance of one substantially different from that of the other.
Note also that the runtime complexities of these methods both require a number of SVD or other decompositions which scales as $N^2$.
Hence in both runtime and performance, at least on this test, these methods are comparable.

\begin{table}
\begin{tabular}{lrrr}
$\alpha$ & $k_{{\rm eff}, \epsilon=10^{-8}}^{{\rm Greedy}}$ & $k_{{\rm eff}, \epsilon=10^{-9}}^{{\rm Greedy}}$ & $k_{{\rm eff}, \epsilon=10^{-8}}^{{\rm Adaptive}}$\\
\hline
0 & 2.3 & 2.3 & 2.4 \\
0.25  & 3.4 & 3.4 & 3.8 \\
0.5  & 3.9 & 4.0 & 3.8 \\
0.75  & 5.0 & 5.1 & 4.7 \\
1.0  & 3.4 & 3.6 & 3.9 \\
\hline	
\end{tabular}
\caption{The effective rank $k_{\rm eff}$  for greedy factorizations of the tensor in equation~\eqref{eq:A} are shown for five values of $\alpha$. For comparison the corresponding effective rank of the adaptive method in~\citet{Ballani2014} is also shown.}
\label{tab:compare}	
\end{table}

\section{Conclusions}

High-rank tensors are increasingly common in physical computation, and so there is now a demand for algorithms which can manipulate them efficiently.
I have introduced and tested two new algorithms for identifying good tree decompositions of tensors.
The first is a straightforward brute-force approach which, up to truncation error, to produces the optimal result at the expense of an exponential number of trials.
The second is a greedy algorithm which, while not guaranteed to produce the optimal outcome, performs very well in all cases examined while only requiring a quadratic number of comparisons.
These fill a crucial niche among the building blocks for tensor computations.

\section*{Acknowledgements}

I am grateful to the UK Marshall Commission for financial support as well as to Milo Lin and Ravishankar Sundararaman for helpful discussions regarding this work.
This research used resources of the National Energy Research Scientific Computing Center, a DOE Office of Science User Facility supported by the Office of Science of the U.S. Department of Energy under Contract No. DE-AC02-05CH11231.

\appendix

\section{Software Details}
\label{appen:software}

The software used in this work is available under a GPLv3 license at \url{github.com/adamjermyn/TensorDecomp}, and consists of a collection of Python scripts.
For simplicity the implementation of the greedy algorithm does not currently take advantage of the possibility of reusing comparisons.
The brute force algorithm enumerates all possible trees by first constructing a tree with one internal node and three indices and then iteratively inserting nodes attached to additional indices in place of edges.
These were run on Python v3.6 with NumPy v1.12.1\ \citep{5725236} and the plots were created with Matplotlib v1.5.3\ \citep{4160265} and the ggplot style, though no specific features of these releases were used.

\bibliography{refs}

\begin{thebibliography}{26}%
\makeatletter
\providecommand \@ifxundefined [1]{%
 \@ifx{#1\undefined}
}%
\providecommand \@ifnum [1]{%
 \ifnum #1\expandafter \@firstoftwo
 \else \expandafter \@secondoftwo
 \fi
}%
\providecommand \@ifx [1]{%
 \ifx #1\expandafter \@firstoftwo
 \else \expandafter \@secondoftwo
 \fi
}%
\providecommand \natexlab [1]{#1}%
\providecommand \enquote  [1]{``#1''}%
\providecommand \bibnamefont  [1]{#1}%
\providecommand \bibfnamefont [1]{#1}%
\providecommand \citenamefont [1]{#1}%
\providecommand \href@noop [0]{\@secondoftwo}%
\providecommand \href [0]{\begingroup \@sanitize@url \@href}%
\providecommand \@href[1]{\@@startlink{#1}\@@href}%
\providecommand \@@href[1]{\endgroup#1\@@endlink}%
\providecommand \@sanitize@url [0]{\catcode `\\12\catcode `\$12\catcode
  `\&12\catcode `\#12\catcode `\^12\catcode `\_12\catcode `\%12\relax}%
\providecommand \@@startlink[1]{}%
\providecommand \@@endlink[0]{}%
\providecommand \url  [0]{\begingroup\@sanitize@url \@url }%
\providecommand \@url [1]{\endgroup\@href {#1}{\urlprefix }}%
\providecommand \urlprefix  [0]{URL }%
\providecommand \Eprint [0]{\href }%
\providecommand \doibase [0]{http://dx.doi.org/}%
\providecommand \selectlanguage [0]{\@gobble}%
\providecommand \bibinfo  [0]{\@secondoftwo}%
\providecommand \bibfield  [0]{\@secondoftwo}%
\providecommand \translation [1]{[#1]}%
\providecommand \BibitemOpen [0]{}%
\providecommand \bibitemStop [0]{}%
\providecommand \bibitemNoStop [0]{.\EOS\space}%
\providecommand \EOS [0]{\spacefactor3000\relax}%
\providecommand \BibitemShut  [1]{\csname bibitem#1\endcsname}%
\let\auto@bib@innerbib\@empty
\bibitem [{\citenamefont {Orús}(2014)}]{ORUS2014117}%
  \BibitemOpen
  \bibfield  {author} {\bibinfo {author} {\bibfnamefont {R.}~\bibnamefont
  {Orús}},\ }\href {\doibase http://dx.doi.org/10.1016/j.aop.2014.06.013}
  {\bibfield  {journal} {\bibinfo  {journal} {Annals of Physics}\ }\textbf
  {\bibinfo {volume} {349}},\ \bibinfo {pages} {117 } (\bibinfo {year}
  {2014})}\BibitemShut {NoStop}%
\bibitem [{\citenamefont {Xie}\ \emph {et~al.}(2009)\citenamefont {Xie},
  \citenamefont {Jiang}, \citenamefont {Chen}, \citenamefont {Weng},\ and\
  \citenamefont {Xiang}}]{PhysRevLett.103.160601}%
  \BibitemOpen
  \bibfield  {author} {\bibinfo {author} {\bibfnamefont {Z.~Y.}\ \bibnamefont
  {Xie}}, \bibinfo {author} {\bibfnamefont {H.~C.}\ \bibnamefont {Jiang}},
  \bibinfo {author} {\bibfnamefont {Q.~N.}\ \bibnamefont {Chen}}, \bibinfo
  {author} {\bibfnamefont {Z.~Y.}\ \bibnamefont {Weng}}, \ and\ \bibinfo
  {author} {\bibfnamefont {T.}~\bibnamefont {Xiang}},\ }\href {\doibase
  10.1103/PhysRevLett.103.160601} {\bibfield  {journal} {\bibinfo  {journal}
  {Phys. Rev. Lett.}\ }\textbf {\bibinfo {volume} {103}},\ \bibinfo {pages}
  {160601} (\bibinfo {year} {2009})}\BibitemShut {NoStop}%
\bibitem [{\citenamefont {Bachmayr}\ \emph {et~al.}(2016)\citenamefont
  {Bachmayr}, \citenamefont {Schneider},\ and\ \citenamefont
  {Uschmajew}}]{Bachmayr:2016:TNH:3027165.3027175}%
  \BibitemOpen
  \bibfield  {author} {\bibinfo {author} {\bibfnamefont {M.}~\bibnamefont
  {Bachmayr}}, \bibinfo {author} {\bibfnamefont {R.}~\bibnamefont {Schneider}},
  \ and\ \bibinfo {author} {\bibfnamefont {A.}~\bibnamefont {Uschmajew}},\
  }\href {\doibase 10.1007/s10208-016-9317-9} {\bibfield  {journal} {\bibinfo
  {journal} {Found. Comput. Math.}\ }\textbf {\bibinfo {volume} {16}},\
  \bibinfo {pages} {1423} (\bibinfo {year} {2016})}\BibitemShut {NoStop}%
\bibitem [{\citenamefont {Evenbly}\ and\ \citenamefont
  {Vidal}(2011)}]{Evenbly2011}%
  \BibitemOpen
  \bibfield  {author} {\bibinfo {author} {\bibfnamefont {G.}~\bibnamefont
  {Evenbly}}\ and\ \bibinfo {author} {\bibfnamefont {G.}~\bibnamefont
  {Vidal}},\ }\href {\doibase 10.1007/s10955-011-0237-4} {\bibfield  {journal}
  {\bibinfo  {journal} {Journal of Statistical Physics}\ }\textbf {\bibinfo
  {volume} {145}},\ \bibinfo {pages} {891} (\bibinfo {year}
  {2011})}\BibitemShut {NoStop}%
\bibitem [{\citenamefont {Markov}\ and\ \citenamefont
  {Shi}(2008)}]{doi:10.1137/050644756}%
  \BibitemOpen
  \bibfield  {author} {\bibinfo {author} {\bibfnamefont {I.~L.}\ \bibnamefont
  {Markov}}\ and\ \bibinfo {author} {\bibfnamefont {Y.}~\bibnamefont {Shi}},\
  }\href {\doibase 10.1137/050644756} {\bibfield  {journal} {\bibinfo
  {journal} {SIAM Journal on Computing}\ }\textbf {\bibinfo {volume} {38}},\
  \bibinfo {pages} {963} (\bibinfo {year} {2008})},\ \Eprint
  {http://arxiv.org/abs/https://doi.org/10.1137/050644756}
  {https://doi.org/10.1137/050644756} \BibitemShut {NoStop}%
\bibitem [{\citenamefont {Bal}\ \emph {et~al.}(2017)\citenamefont {Bal},
  \citenamefont {Mari\"en}, \citenamefont {Haegeman},\ and\ \citenamefont
  {Verstraete}}]{PhysRevLett.118.250602}%
  \BibitemOpen
  \bibfield  {author} {\bibinfo {author} {\bibfnamefont {M.}~\bibnamefont
  {Bal}}, \bibinfo {author} {\bibfnamefont {M.}~\bibnamefont {Mari\"en}},
  \bibinfo {author} {\bibfnamefont {J.}~\bibnamefont {Haegeman}}, \ and\
  \bibinfo {author} {\bibfnamefont {F.}~\bibnamefont {Verstraete}},\ }\href
  {\doibase 10.1103/PhysRevLett.118.250602} {\bibfield  {journal} {\bibinfo
  {journal} {Phys. Rev. Lett.}\ }\textbf {\bibinfo {volume} {118}},\ \bibinfo
  {pages} {250602} (\bibinfo {year} {2017})}\BibitemShut {NoStop}%
\bibitem [{\citenamefont {Evenbly}(2017)}]{PhysRevB.95.045117}%
  \BibitemOpen
  \bibfield  {author} {\bibinfo {author} {\bibfnamefont {G.}~\bibnamefont
  {Evenbly}},\ }\href {\doibase 10.1103/PhysRevB.95.045117} {\bibfield
  {journal} {\bibinfo  {journal} {Phys. Rev. B}\ }\textbf {\bibinfo {volume}
  {95}},\ \bibinfo {pages} {045117} (\bibinfo {year} {2017})}\BibitemShut
  {NoStop}%
\bibitem [{\citenamefont {Yang}\ \emph {et~al.}(2017)\citenamefont {Yang},
  \citenamefont {Gu},\ and\ \citenamefont {Wen}}]{PhysRevLett.118.110504}%
  \BibitemOpen
  \bibfield  {author} {\bibinfo {author} {\bibfnamefont {S.}~\bibnamefont
  {Yang}}, \bibinfo {author} {\bibfnamefont {Z.-C.}\ \bibnamefont {Gu}}, \ and\
  \bibinfo {author} {\bibfnamefont {X.-G.}\ \bibnamefont {Wen}},\ }\href
  {\doibase 10.1103/PhysRevLett.118.110504} {\bibfield  {journal} {\bibinfo
  {journal} {Phys. Rev. Lett.}\ }\textbf {\bibinfo {volume} {118}},\ \bibinfo
  {pages} {110504} (\bibinfo {year} {2017})}\BibitemShut {NoStop}%
\bibitem [{\citenamefont {Evenbly}\ and\ \citenamefont
  {Pfeifer}(2014)}]{PhysRevB.89.245118}%
  \BibitemOpen
  \bibfield  {author} {\bibinfo {author} {\bibfnamefont {G.}~\bibnamefont
  {Evenbly}}\ and\ \bibinfo {author} {\bibfnamefont {R.~N.~C.}\ \bibnamefont
  {Pfeifer}},\ }\href {\doibase 10.1103/PhysRevB.89.245118} {\bibfield
  {journal} {\bibinfo  {journal} {Phys. Rev. B}\ }\textbf {\bibinfo {volume}
  {89}},\ \bibinfo {pages} {245118} (\bibinfo {year} {2014})}\BibitemShut
  {NoStop}%
\bibitem [{\citenamefont {Pfeifer}\ \emph {et~al.}(2014)\citenamefont
  {Pfeifer}, \citenamefont {Haegeman},\ and\ \citenamefont
  {Verstraete}}]{PhysRevE.90.033315}%
  \BibitemOpen
  \bibfield  {author} {\bibinfo {author} {\bibfnamefont {R.~N.~C.}\
  \bibnamefont {Pfeifer}}, \bibinfo {author} {\bibfnamefont {J.}~\bibnamefont
  {Haegeman}}, \ and\ \bibinfo {author} {\bibfnamefont {F.}~\bibnamefont
  {Verstraete}},\ }\href {\doibase 10.1103/PhysRevE.90.033315} {\bibfield
  {journal} {\bibinfo  {journal} {Phys. Rev. E}\ }\textbf {\bibinfo {volume}
  {90}},\ \bibinfo {pages} {033315} (\bibinfo {year} {2014})}\BibitemShut
  {NoStop}%
\bibitem [{\citenamefont {Kolda}\ and\ \citenamefont
  {Bader}(2009)}]{doi:10.1137/07070111X}%
  \BibitemOpen
  \bibfield  {author} {\bibinfo {author} {\bibfnamefont {T.~G.}\ \bibnamefont
  {Kolda}}\ and\ \bibinfo {author} {\bibfnamefont {B.~W.}\ \bibnamefont
  {Bader}},\ }\href {\doibase 10.1137/07070111X} {\bibfield  {journal}
  {\bibinfo  {journal} {SIAM Review}\ }\textbf {\bibinfo {volume} {51}},\
  \bibinfo {pages} {455} (\bibinfo {year} {2009})},\ \Eprint
  {http://arxiv.org/abs/https://doi.org/10.1137/07070111X}
  {https://doi.org/10.1137/07070111X} \BibitemShut {NoStop}%
\bibitem [{\citenamefont {{Ying}}(2016)}]{2016arXiv160700050Y}%
  \BibitemOpen
  \bibfield  {author} {\bibinfo {author} {\bibfnamefont {L.}~\bibnamefont
  {{Ying}}},\ }\href@noop {} {\bibfield  {journal} {\bibinfo  {journal} {ArXiv
  e-prints}\ } (\bibinfo {year} {2016})},\ \Eprint
  {http://arxiv.org/abs/1607.00050} {arXiv:1607.00050 [math.NA]} \BibitemShut
  {NoStop}%
\bibitem [{\citenamefont {Vidal}(2008)}]{PhysRevLett.101.110501}%
  \BibitemOpen
  \bibfield  {author} {\bibinfo {author} {\bibfnamefont {G.}~\bibnamefont
  {Vidal}},\ }\href {\doibase 10.1103/PhysRevLett.101.110501} {\bibfield
  {journal} {\bibinfo  {journal} {Phys. Rev. Lett.}\ }\textbf {\bibinfo
  {volume} {101}},\ \bibinfo {pages} {110501} (\bibinfo {year}
  {2008})}\BibitemShut {NoStop}%
\bibitem [{\citenamefont {Nakatani}\ and\ \citenamefont
  {Chan}(2013)}]{doi:10.1063/1.4798639}%
  \BibitemOpen
  \bibfield  {author} {\bibinfo {author} {\bibfnamefont {N.}~\bibnamefont
  {Nakatani}}\ and\ \bibinfo {author} {\bibfnamefont {G.~K.-L.}\ \bibnamefont
  {Chan}},\ }\href {\doibase 10.1063/1.4798639} {\bibfield  {journal} {\bibinfo
   {journal} {The Journal of Chemical Physics}\ }\textbf {\bibinfo {volume}
  {138}},\ \bibinfo {pages} {134113} (\bibinfo {year} {2013})},\ \Eprint
  {http://arxiv.org/abs/http://dx.doi.org/10.1063/1.4798639}
  {http://dx.doi.org/10.1063/1.4798639} \BibitemShut {NoStop}%
\bibitem [{\citenamefont {Hackbusch}\ and\ \citenamefont
  {K{\"u}hn}(2009)}]{Hackbusch2009}%
  \BibitemOpen
  \bibfield  {author} {\bibinfo {author} {\bibfnamefont {W.}~\bibnamefont
  {Hackbusch}}\ and\ \bibinfo {author} {\bibfnamefont {S.}~\bibnamefont
  {K{\"u}hn}},\ }\href {\doibase 10.1007/s00041-009-9094-9} {\bibfield
  {journal} {\bibinfo  {journal} {Journal of Fourier Analysis and
  Applications}\ }\textbf {\bibinfo {volume} {15}},\ \bibinfo {pages} {706}
  (\bibinfo {year} {2009})}\BibitemShut {NoStop}%
\bibitem [{\citenamefont {Ballani}\ \emph {et~al.}(2013)\citenamefont
  {Ballani}, \citenamefont {Grasedyck},\ and\ \citenamefont
  {Kluge}}]{BALLANI2013639}%
  \BibitemOpen
  \bibfield  {author} {\bibinfo {author} {\bibfnamefont {J.}~\bibnamefont
  {Ballani}}, \bibinfo {author} {\bibfnamefont {L.}~\bibnamefont {Grasedyck}},
  \ and\ \bibinfo {author} {\bibfnamefont {M.}~\bibnamefont {Kluge}},\ }\href
  {\doibase http://dx.doi.org/10.1016/j.laa.2011.08.010} {\bibfield  {journal}
  {\bibinfo  {journal} {Linear Algebra and its Applications}\ }\textbf
  {\bibinfo {volume} {438}},\ \bibinfo {pages} {639 } (\bibinfo {year}
  {2013})},\ \bibinfo {note} {tensors and Multilinear Algebra}\BibitemShut
  {NoStop}%
\bibitem [{\citenamefont {Ballani}\ and\ \citenamefont
  {Grasedyck}(2014)}]{Ballani2014}%
  \BibitemOpen
  \bibfield  {author} {\bibinfo {author} {\bibfnamefont {J.}~\bibnamefont
  {Ballani}}\ and\ \bibinfo {author} {\bibfnamefont {L.}~\bibnamefont
  {Grasedyck}},\ }\href {\doibase 10.1137/130926328} {\bibfield  {journal}
  {\bibinfo  {journal} {{SIAM} Journal on Scientific Computing}\ }\textbf
  {\bibinfo {volume} {36}},\ \bibinfo {pages} {A1415} (\bibinfo {year}
  {2014})}\BibitemShut {NoStop}%
\bibitem [{\citenamefont {MISZCZAK}(2011)}]{doi:10.1142/S0129183111016683}%
  \BibitemOpen
  \bibfield  {author} {\bibinfo {author} {\bibfnamefont {J.~A.}\ \bibnamefont
  {MISZCZAK}},\ }\href {\doibase 10.1142/S0129183111016683} {\bibfield
  {journal} {\bibinfo  {journal} {International Journal of Modern Physics C}\
  }\textbf {\bibinfo {volume} {22}},\ \bibinfo {pages} {897} (\bibinfo {year}
  {2011})},\ \Eprint
  {http://arxiv.org/abs/http://www.worldscientific.com/doi/pdf/10.1142/S0129183111016683}
  {http://www.worldscientific.com/doi/pdf/10.1142/S0129183111016683}
  \BibitemShut {NoStop}%
\bibitem [{\citenamefont {Penrose}(1971)}]{Penrose}%
  \BibitemOpen
  \bibfield  {author} {\bibinfo {author} {\bibfnamefont {R.}~\bibnamefont
  {Penrose}},\ }\href@noop {} {\bibfield  {journal} {\bibinfo  {journal}
  {Combinatorial Mathematics and its Applications}\ } (\bibinfo {year}
  {1971})}\BibitemShut {NoStop}%
\bibitem [{\citenamefont {Eckart}\ and\ \citenamefont
  {Young}(1936)}]{Eckart1936}%
  \BibitemOpen
  \bibfield  {author} {\bibinfo {author} {\bibfnamefont {C.}~\bibnamefont
  {Eckart}}\ and\ \bibinfo {author} {\bibfnamefont {G.}~\bibnamefont {Young}},\
  }\href {\doibase 10.1007/BF02288367} {\bibfield  {journal} {\bibinfo
  {journal} {Psychometrika}\ }\textbf {\bibinfo {volume} {1}},\ \bibinfo
  {pages} {211} (\bibinfo {year} {1936})}\BibitemShut {NoStop}%
\bibitem [{\citenamefont {Yamanaka}\ \emph {et~al.}(2009)\citenamefont
  {Yamanaka}, \citenamefont {Otachi},\ and\ \citenamefont
  {Nakano}}]{Yamanaka2009}%
  \BibitemOpen
  \bibfield  {author} {\bibinfo {author} {\bibfnamefont {K.}~\bibnamefont
  {Yamanaka}}, \bibinfo {author} {\bibfnamefont {Y.}~\bibnamefont {Otachi}}, \
  and\ \bibinfo {author} {\bibfnamefont {S.-i.}\ \bibnamefont {Nakano}},\
  }\enquote {\bibinfo {title} {Efficient enumeration of ordered trees with k
  leaves (extended abstract)},}\ in\ \href {\doibase
  10.1007/978-3-642-00202-1_13} {\emph {\bibinfo {booktitle} {WALCOM:
  Algorithms and Computation: Third International Workshop, WALCOM 2009,
  Kolkata, India, February 18-20, 2009. Proceedings}}},\ \bibinfo {editor}
  {edited by\ \bibinfo {editor} {\bibfnamefont {S.}~\bibnamefont {Das}}\ and\
  \bibinfo {editor} {\bibfnamefont {R.}~\bibnamefont {Uehara}}}\ (\bibinfo
  {publisher} {Springer Berlin Heidelberg},\ \bibinfo {address} {Berlin,
  Heidelberg},\ \bibinfo {year} {2009})\ pp.\ \bibinfo {pages}
  {141--150}\BibitemShut {NoStop}%
\bibitem [{\citenamefont {Zaks}\ and\ \citenamefont
  {Richards}(1979)}]{doi:10.1137/0208006}%
  \BibitemOpen
  \bibfield  {author} {\bibinfo {author} {\bibfnamefont {S.}~\bibnamefont
  {Zaks}}\ and\ \bibinfo {author} {\bibfnamefont {D.}~\bibnamefont
  {Richards}},\ }\href {\doibase 10.1137/0208006} {\bibfield  {journal}
  {\bibinfo  {journal} {SIAM Journal on Computing}\ }\textbf {\bibinfo {volume}
  {8}},\ \bibinfo {pages} {73} (\bibinfo {year} {1979})},\ \Eprint
  {http://arxiv.org/abs/https://doi.org/10.1137/0208006}
  {https://doi.org/10.1137/0208006} \BibitemShut {NoStop}%
\bibitem [{\citenamefont {Ebrahim}(2009)}]{doi:10.1093/tropej/fmn004}%
  \BibitemOpen
  \bibfield  {author} {\bibinfo {author} {\bibfnamefont {G.~J.}\ \bibnamefont
  {Ebrahim}},\ }\href {\doibase 10.1093/tropej/fmn004} {\bibfield  {journal}
  {\bibinfo  {journal} {Journal of Tropical Pediatrics}\ }\textbf {\bibinfo
  {volume} {55}},\ \bibinfo {pages} {67} (\bibinfo {year} {2009})}\BibitemShut
  {NoStop}%
\bibitem [{\citenamefont {Molina-Vilaplana}\ and\ \citenamefont
  {Prior}(2014)}]{MolinaVilaplana2014}%
  \BibitemOpen
  \bibfield  {author} {\bibinfo {author} {\bibfnamefont {J.}~\bibnamefont
  {Molina-Vilaplana}}\ and\ \bibinfo {author} {\bibfnamefont {J.}~\bibnamefont
  {Prior}},\ }\href {\doibase 10.1007/s10714-014-1823-y} {\bibfield  {journal}
  {\bibinfo  {journal} {General Relativity and Gravitation}\ }\textbf {\bibinfo
  {volume} {46}},\ \bibinfo {pages} {1823} (\bibinfo {year}
  {2014})}\BibitemShut {NoStop}%
\bibitem [{\citenamefont {van~der Walt}\ \emph {et~al.}(2011)\citenamefont
  {van~der Walt}, \citenamefont {Colbert},\ and\ \citenamefont
  {Varoquaux}}]{5725236}%
  \BibitemOpen
  \bibfield  {author} {\bibinfo {author} {\bibfnamefont {S.}~\bibnamefont
  {van~der Walt}}, \bibinfo {author} {\bibfnamefont {S.~C.}\ \bibnamefont
  {Colbert}}, \ and\ \bibinfo {author} {\bibfnamefont {G.}~\bibnamefont
  {Varoquaux}},\ }\href {\doibase 10.1109/MCSE.2011.37} {\bibfield  {journal}
  {\bibinfo  {journal} {Computing in Science Engineering}\ }\textbf {\bibinfo
  {volume} {13}},\ \bibinfo {pages} {22} (\bibinfo {year} {2011})}\BibitemShut
  {NoStop}%
\bibitem [{\citenamefont {Hunter}(2007)}]{4160265}%
  \BibitemOpen
  \bibfield  {author} {\bibinfo {author} {\bibfnamefont {J.~D.}\ \bibnamefont
  {Hunter}},\ }\href {\doibase 10.1109/MCSE.2007.55} {\bibfield  {journal}
  {\bibinfo  {journal} {Computing in Science Engineering}\ }\textbf {\bibinfo
  {volume} {9}},\ \bibinfo {pages} {90} (\bibinfo {year} {2007})}\BibitemShut
  {NoStop}%
\end{thebibliography}%

\end{document}